\documentclass[11pt,letter]{iopart}

\usepackage{epsfig,amssymb}

\newcommand{\ZG}{ZnGa$_2$O$_4$} 
\newcommand{\F}{Fe$_3$O$_4$}
\newcommand{\ZGF}{[ZnGa$_2$O$_4$]$_{1-x}$[Fe$_3$O$_4$]$_x$}

\begin{document}

\title{Dilute ferrimagnetic semiconductors in Fe-substituted spinel 
\ZG\/} 

\author{A. S. Risbud\dag, R. Seshadri\dag, J. Ensling\ddag, and C. Felser\ddag}

\address{\dag Materials Department and Materials Research Laboratory\\
         University of California, Santa Barbara CA 93106\\
         seshadri@mrl.ucsb.edu \ \ http://www.mrl.ucsb.edu/$\sim$seshadri}

\address{\ddag Institut f\"ur Anorganische Chemie und Analytische Chemie\\
Johannes Gutenberg-Universit\"at, Staudinger Weg 9, 55099 Mainz}

\begin{abstract}
Solid solutions of nominal composition \ZGF, of the semiconducting spinel \ZG\/
with the ferrimagnetic spinel \F\/ have been prepared with $x$ = 0.05, 0.10, 
and 0.15.  All samples show evidence for long-range magnetic
ordering with ferromagnetic hysteresis at low temperatures. Magnetization as
a function of field for the $x$ = 0.15 sample is \textsf{S}-shaped
at temperatures as high as 200 K. M\"ossbauer spectroscopy on the $x$ = 0.15 
sample confirms the presence of Fe$^{3+}$, and spontaneous magnetization at 
4.2 K.  The magnetic behavior is obtained without greatly affecting the 
semiconducting properties of the host; diffuse reflectance optical 
spectroscopy indicates that Fe substitution up to $x$ = 0.15 does not affect 
the position of the band edge absorption. These promising results motivate the 
possibility of dilute ferrimagnetic semiconductors which do not require 
carrier mediation of the magnetic moment.
\end{abstract}

\pacs{75.50.Pp, 
      75.50.Gg, 
      78.20.-e 
      }

\maketitle

\section{Introduction}

The burgeoning field of spintronics \cite{Awschalom_SA2002,Dassarma_AS2001}
has created demands for entirely new classes of materials. One such materials
class combines both semiconducting and ferromagnetic properties. 
Magnetic semiconductors can help in the efficient injection of spin from a 
ferromagnetic contact to a semiconductor.\cite{Rashba_PRB2000} 
The substitution of magnetic ions such as Mn$^{3+}$ in traditional 
semiconducting materials such as GaAs is a well-established route 
to making ferromagnetic semiconductors.\cite{Ohno_SCIENCE1998,Ohno_JMMM1999} 
However Mn-substituted GaAs as well as digital heterostructures 
such as GaAs/0.5ML-Mn \cite{Kawakami_APL2000} have maximum Curie temperatures 
of around 100 K or less.  In the ideal case,
we desire magnetic semiconductors with a magnetic transition above room
temperature. The ability to have both $n$- and $p$-type doping, a long
spin-relaxation lifetime, and large carrier spin-polarization are other
desirable features. Proposed candidate materials are strongly hole-doped 
wide band gap wurtzite semiconductors GaN and ZnO, with magnetic transition
metal \textit{t}M substituents on the cation 
sites.\cite{Dietl_SCIENCE2000,Pearton_JAP2003} 
A number of recent studies on bulk and thin-film samples have ensued 
which suggest that ZnO:\textit{t}M with \textit{t}M = Co 
\cite{Ueda_APL2001,Kim_PhysicaB2003,Lim_SSC2003} and with 
\textit{t}M = Mn \cite{Sharma_NatMat2003,Yeo_APL2004}
are ferromagnetic at room temperature, whilst others find spin-glass behavior.
\cite{Fukumura_APL2001,Kolesnik_JSupercond2002}

Results of our prior work on bulk samples of ZnO:\textit{t}M systems suggest a
complete absence of any magnetic ordering in well-characterized bulk 
samples.\cite{Risbud_PRB2003,Lawes_XXX2004} Studies by 
Kolesnik \textit{et al.\/} and coworkers confirm these 
findings.\cite{Kolesnik_JAP} Careful density functional 
calculations by Spaldin \cite{Spaldin_PRB2004} also point to difficulties in
inducing ferromagnetism by transition metal substitution in ZnO. We find that 
while magnetic 
susceptibility indicates strong near-neighbor coupling in ZnO:\textit{t}M 
(\textit{t}M = Mn and Co), the mean-field coupling is exceedingly 
weak.\cite{Lawes_XXX2004} It is the mean field term that would dictate 
magnetic ordering.  It is useful to temper expectation with the
following observation: the few known ferromagnetic oxide semiconductors
have rather low $T_C$s: BiMnO$_3$ ($T_C$ = 105 K), \cite{Chiba_JSSC1997}
SeCuO$_3$ ($T_C$ = 26 K), \cite{Kohn_JSSC1976} 
YTiO$_3$ ($T_C$ =  29 K), \cite{Garrett_MRB1981}  
EuO ($T_C$ = 79 K).\cite{Matthias_PRL1961}

\begin{figure} 
\centering \epsfig{file=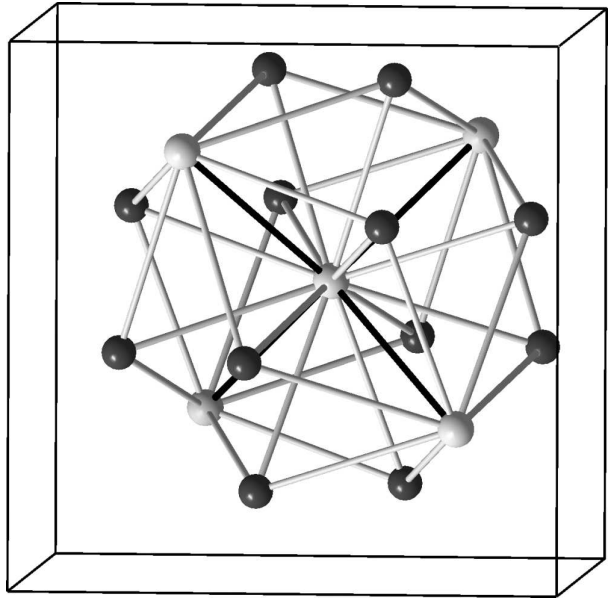, width=8cm} 
\caption{Cation cluster centered around an tetrahedral A atom (light gray 
spheres) in the spinel crystal structure. The dark gray spheres are the 
octahedral B atoms. Within a 3.7 \AA\/ distance, each octahedral Ga atom in
\ZG\/ has 12 tetrahedral Zn neighbors and 4 Ga neighbors.}
\label{fig:Structure} \end{figure}

A promising alternative is to induce \textit{ferrimagnetism\/} in a wide
band gap semiconductor with \textit{two\/} host cation sites: this would
exploit the more natural tendency of spins in insulating  oxides to anti-align 
and would require no conduction electrons. A natural structure for 
ferrimagnetism is spinel, AB$_2$O$_4$, where spins
in the tetrahedral (A) and  octahedral (B) cation sites are usually
anti-aligned with respect to each other; in addition, with two B sites 
for every A site, the possibility of a net ferrimagnetic moment exists. 
Figure~\ref{fig:Structure} shows the cation network centered around a single
tetrahedral A cation in the spinel structure. The very high effective
cation coordination in this structure type would mean that even at rather
low concentrations of substituent transition metal ions, extensive magnetic 
coupling can be expected. In this contribution, we have chosen spinel \ZG\/ 
with a direct band gap of 4.1 eV as host.\cite{Sampath_JACerS1998} 
Both \ZG\/ as well as ZnO can be grown epitaxially on spinel MgAl$_2$O$_4$ 
substrates  \cite{Andeen_JCrystG2003, Loeffler_JMR2004}, 
suggesting that the title compound can be incorporated in devices.

By preparing nominal solid solutions \ZGF\/ ($0 \le x \le 0.15$), we have been
able to induce magnetism, with strongly hysteretic behavior
for the $x$ = 0.15 sample at 5 K. M\"ossbauer spectra acquired at 293 K and 
4.2 K on the $x$ = 0.15 sample suggests spontaneous magnetization at 4.2 K.
The data also reveal the complete absence of Fe$^{2+}$ in the sample, which
supports our view of dilute ferrimagnetism rather than magnetism arising from 
a spurious source such as magnetite Fe$_3$O$_4$ nanoparticles in a \ZG\/ 
matrix. Diffuse reflectance UV/Vis spectroscopy reveals that the direct band 
gap of \ZG\/ is unaffected by substitution. 

\section{Experimental}

As in previous work on polycrystalline Zn$_{1-x}$M$_x$O materials 
\cite{Risbud_PRB2003,Lawes_XXX2004} precursor oxalates 
Zn$_{1-x}$Fe$_x$(C$_2$O$_4$)$\cdot$2H$_2$O were prepared with 
$x$ = 0.00, 0.02, 0.05, 0.10, and 0.15 by precipitation from aqueous solution. 
Polycrystalline samples with the nominal compositions \ZGF\/ 
($0 \le x \le 0.15)$ were made by grinding together the corresponding oxalate 
with appropriate amounts of Ga$_2$O$_3$ and 
decomposing in air at 1473 K for 18 hrs., with an intermediate regrinding 
step. For $x$ = 0.00, the powder is white; as $x$ increases, the powder colors 
range from beige to auburn. Beyond $x$ = 0.15, samples require longer sintering
times, suggesting a possible solubility limit in the vicinity. Step-scanned 
X-ray diffraction data on the powders were collected on a Scintag X-2 
diffractometer operated in the $\theta$-$2\theta$ geometry. Transmission 
electron microscopy of the $x$ = 0.15 sample, in conjunction with energy 
dispersive X-ray spectroscopic (EDX) analysis was carried out on a 
JEOL JEM 2010 microscope. 
The sample powder was dispersed from a solvent on to a carbon-coated copper 
grid for TEM studies. UV/Vis diffuse reflectance spectra were acquired 
on powders sprinkled on scotch tape. Magnetization data were collected on a 
Quantum Design MPMS 5XL SQUID magnetometer operated between 2 K and 400 K.
M\"ossbauer spectra were collected at 293 K and 4.2 K using a 
constant-acceleration spectrometer equipped with a 1024 channel analyzer and
operated in the time scale mode. The $\gamma$ source was 25 mCi $^{57}$Co/Rh.
The spectra were analyzed using the computer program 
\textsc{effino}.\cite{EFFINO} 

\section{Results and discussion}

\begin{figure} 
\centering \epsfig{file=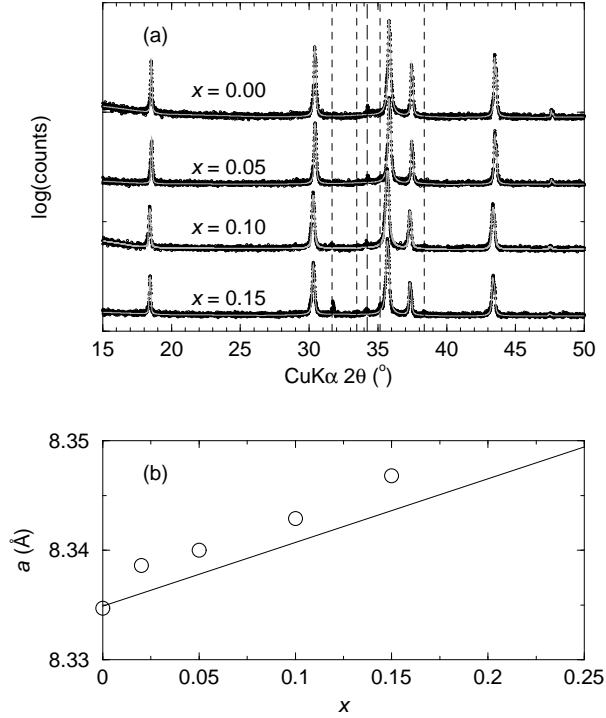, width=8cm} 
\caption{(a) Powder X-ray diffraction patterns of the nominal samples
\ZGF\/ ($x$ = 0, 0.05, 0.10, and 0.15) displayed on a log scale. 
Points are data and the line is the Rietveld fit to the spinel crystal 
structure. The four dashed vertical lines indicate a very small impurity of 
monoclinic Ga$_2$O$_3$. The long dashed line at 34.2$^{\circ}$ 2$\theta$,
is common to all the samples, is an unidentified non-magnetic impurity. 
(b) Circles display the evolution of the spinel
lattice parameter with $x$ in the region $0 \le x \le 0.15$. 
The line connects the published cell parameter of spinel \ZG\/ at $x$ = 0 with
the published cell parameter of spinel \F\/ at $x$ = 1.}
\label{fig:X-ray} \end{figure}

X-ray diffraction data, plotted on a log scale in Figure~\ref{fig:X-ray}(a), 
indicate a very small as-yet unidentified impurity at 34.2$^{\circ}$ 2$\theta$
in all the samples, including the non-magnetic host \ZG.
In addition, the $x$ = 0.15 sample has a small impurity of monoclinic 
Ga$_2$O$_3$ (PDF Card 41-1103). Other than these clearly non-magnetic 
impurities, the samples are clean showing only the spinel phase. The 
diffraction profiles could be satisfactorily fit to the spinel crystal 
structure by the Rietveld method using the \textsc{xnd} 
program.\cite{Berar_xnd} 
For all the samples, reliability factor ($R_{\rm Bragg}$) was less than 7\%.
The evolution of the $a$ cell parameter with $x$ is shown in 
Figure~\ref{fig:X-ray}(b)  and is consistent with substitution of Fe on the 
spinel \ZG\/ lattice.  The line connects published cell parameters of 
\ZG\/ \cite{Josties_NJMM1995} and \F.\cite{Fleet_ActaCB1982} Under the 
conditions of our preparation, Fe$_3$O$_4$ is stable under the heat
treatment conditions which we have employed,\cite{Sundman_JPhaseEq1991} 
so we were initially confident in describing the Fe substitution as a solid 
solution \ZGF. However, as we shall describe presently, M\"ossbauer spectra 
indicate a complete absence of Fe$^{2+}$ in the $x$ = 0.15 sample. 
We do see a small broadening of the X-ray profiles with substitution 
characteristic of the increased number of elements in the crystal. 
The X-ray form factors of Fe, Zn and Ga are too similar to one-another
for us to directly refine the relative amounts of these elements in the
spinel crystal structure.

TEM/EDX analysis of the $x$ = 0.15 sample systematically indicated
that the Fe:Zn:Ga atomic ratio was close to 1:3:6 rather than the ratio 
calculated from the nominal composition which is 1:1.9:3.8, or
nearly 1:2:4. More importantly, the samples are homogeneous on the length 
scale of the EDX spot size (approximately 3 nm), 
consistently displaying the same ratios over several spots. 

\begin{figure} 
~\\
\centering \epsfig{file=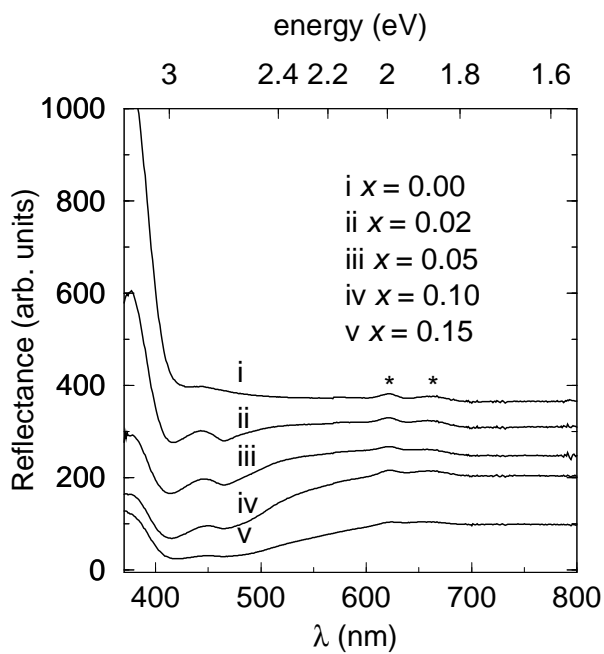, width=8cm} 
\caption{Diffuse reflectance UV/Vis spectra of \ZGF\/ for the different $x$ in 
the near UV and visible region of the spectrum. Data have been offset along
the reflectance axis for clarity. The asterisks are associated with the 
absorption of the Scotch tape.}
\label{fig:Optical} 
\end{figure}

Across the substitution series, the semiconducting nature of \ZG\/ is retained
as confirmed by UV/Vis diffuse-reflectance spectroscopy. Diffuse reflectance
UV/Vis spectra across the series are shown in Figure~\ref{fig:Optical}. 
\ZG\/ is a direct band gap semiconductor,\cite{Sampath_JACerS1998} as 
confirmed by the sharp band-edge absorption. In our samples we find the
band edge is at 3.3 eV rather than the reported value of 4.1 eV. 
However, it is well known\cite{Sampath_JACerS1998} that \ZG\/ displays
a very sensitive dependence of the optical absorption edge on composition,
and a very small concentration of cation vacancies are sufficient to
shift the edge to the red. Substitution by iron leaves the band-edge 
absorption unchanged, though new features associated with atomic transitions 
arise in the visible region. The relative rise of these new features in the 
visible region of the spectrum results in an effective \textit{relative\/} 
decrease in the intensity of the band-edge absorption.

\begin{figure} 
~\\
\centering \epsfig{file=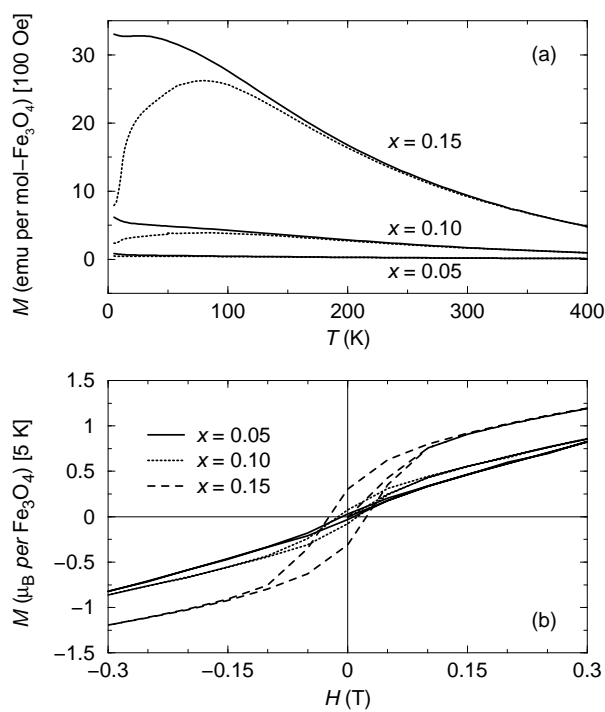, width=8cm} 
\caption{(a) Magnetization under a 100 Oe field as a function of temperature 
for \ZGF\/ with $x$ = 0.05, 0.10, and 0.15. Data were collected upon warming
after cooling under zero field (broken lines) and after cooling under a 100 Oe
field (solid lines). (b) 5 K magnetization of the three samples.}
\label{fig:Magnetism1} 
\end{figure}

Figure~\ref{fig:Magnetism1}(a) shows the temperature dependence of the 
magnetization of three samples, $x$ = 0.05, 0.10, and 0.15. The magnetization
has been scaled by the mole number of Fe$_3$O$_4$; implying that in the 
absence of long-range interactions, traces for the three samples
should collapse on a single curve. We do not observe this even at 400 K.
Plots of $1/M$ as a function of $T$ are not linear for any of the samples below
400 K, also in accord with the existence of long-range interactions.
All three samples show a separation of the ZFC and FC traces. This is clearly
visible for $x$ = 0.10 and $x$ = 0.15 at about 100 K. At 5 K, all three 
samples display hysteretic behavior as seen in Figure~\ref{fig:Magnetism1}(b). 
None of the samples show magnetic saturation indicating that a certain 
fraction of the spins are ``free'' and are not involved in the ordering. 
If we assume that Fe substitutes on \ZG\/ as an alloy with \F, then the 
saturation magnetization should be 4 $\mu_{\rm B}$ \textit{per\/} \F. 
If we take the minimum magnetic saturation value to be the point where the 
hysteresis loop closes, then for $x$ = 0.15, this value is about 
1 $\mu_{\rm B}$ \textit{per\/} \F\/ at 5 K. Therefore, to a good approximation,
25\% of the substituted Fe completely participate in the bulk ferrimagnetism.

\begin{figure} 
~\\
\centering \epsfig{file=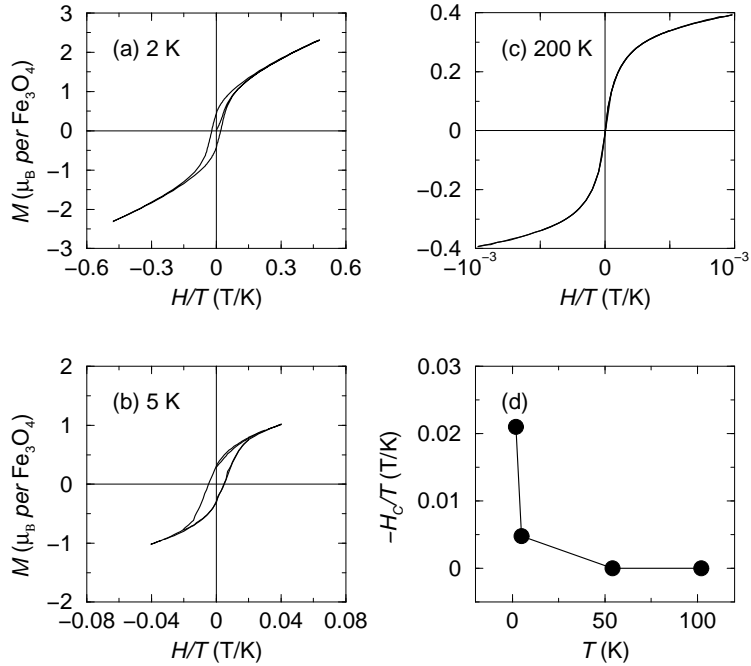, width=10cm} 
\caption{(a-c) Magnetization as a function of $H/T$ of \ZGF\ with $x$ = 0.15 
at different temperatures (indicated). (d) Coercive field scaled by temperature
at different temperatures for the $x$ = 0.15 sample.}
\label{fig:Magnetism2} \end{figure}

Figure~\ref{fig:Magnetism2} displays magnetization for $x$ = 0.15 as a 
function of the scaled field ($H/T$) at three different temperatures, 
(a) 2 K, (b) 5 K, and (c) 200 K. At the two lower temperatures, hysteretic 
behavior is observed. The hysteresis is lost by about 100 K, but even at 
200 K, magnetization as a function of field indicates long range ordering. 
The nature of the magnetization loops would suggest superparamagnetic 
behavior with a blocking temperature of around 100 K. However, 
superparamagnetism would imply that plots of $M$ as a function of $H/T$ would
collapse on to a single \textsf{S}-shaped trace above the blocking temperature.
We do not find this. Figure~\ref{fig:Magnetism2}(d) shows the temperature 
dependence of the magnetic coercivity of $x$ = 0.15, which grows almost 
exponentially as the temperature is lowered. At 5 K, the coercivity is about
250 Oe, to be compared with a value of 420 Oe for 
Fe$_3$O$_4$.\cite{Smirnov_EPSL2002} 

\begin{figure} 
\centering \epsfig{file=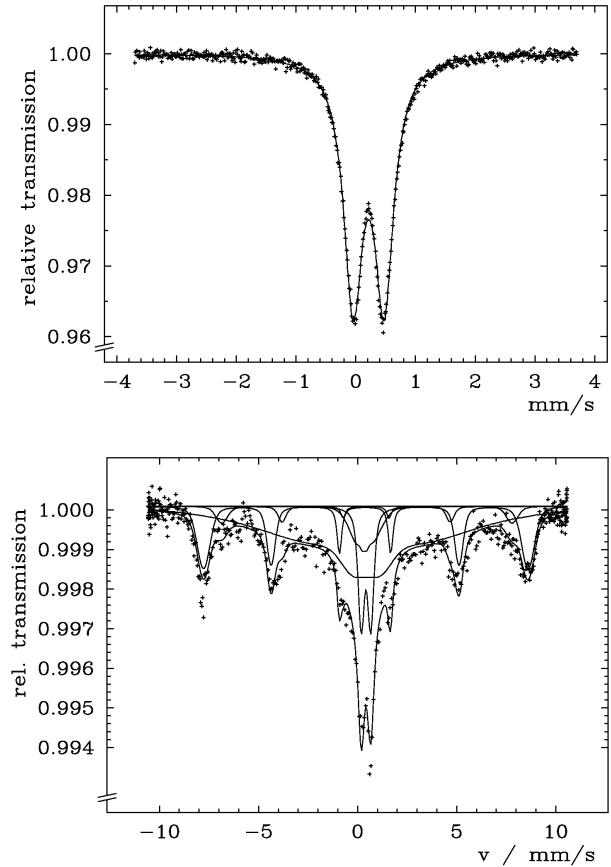, width=8cm} 
\caption{M\"ossbauer spectra acquired at 293 K (upper panel) and 4.2 K (lower
panel) of the nominal \ZGF\/ sample with $x$ = 0.15.}
\label{fig:Moess}\end{figure}

M\"ossbauer spectra were acquired at room temperature (293 K) on $x$ = 0.05
sample and the $x$ = 0.15 sample. For both samples, the isomer shift
with respect to Fe metal was 0.32 mm s$^{-1}$ and the quadrupolar splitting 
was 0.52. Neither sample showed, at 293 K, the 6-finger pattern characteristic
of magnetic ordering. At 4.2 K, the $x$ = 0.15 sample displayed both 
a doublet corresponding to paramagnetic Fe$^{3+}$ and a 6-finger pattern
corresponding to magnetically ordered Fe$^{3+}$. The internal field is between
250 and  510 Oe, consistent with the magnetization data. From the 
M\"ossbauer fit to the para and ferromagnetic phases at 4.2 K, a relative
ratio of 10\% (paramagnetic) to 90\% (ferromagnetic) was obtained. 
M\"ossbauer data and the fit to different components, acquired on the 
$x$ = 0.15 sample are shown in Figure~\ref{fig:Moess}. 
The relative distribution of Fe$^{3+}$ on octahedral and tetrahedral sites 
could not be obtained from the present set of experiments. Experiments are
planned in future with labeled $^{57}$Fe in order to obtain higher resolution
data that will allow the Fe ion site selection to be probed.

The title compounds have random substitution
of ferric ions, presumably on both the tetrahedral and octahedral sites in
the semiconducting host spinel structure. Since there are twice as many 
octahedral sites as there are tetrahedral sites, if the spins behave as they
do in bulk \F, ferromagnetism can ensue. We believe this to be at the origin
of the observed magnetic behavior. Further M\"ossbauer studies as well as 
powder neutron diffraction measurements are planned in order to better 
characterize the title compounds. Our results suggest that dilute 
\textit{ferri}magnetic semiconductors are worthy objects of study.

\ack We thank Gavin Lawes for useful discussions and Dan Cohen for help with 
the optical measurements, and Ombretta Masala for help with TEM studies.
A. S. R. is supported by the National Science Foundation
IGERT program under the award DGE-9987618. We gratefully acknowledge support
from a CARE Award (SBB-0304A) of the University of California/Los Alamos 
National Laboratory. The work made use of facilities supported by the MRL 
program of the National Science Foundation under the Award No. DMR00-80034.

\pagebreak

\clearpage


\begin{thebibliography}{99}

\bibitem{Awschalom_SA2002} Awschalom D D, Flatte M E and Samarth N 2002 
\textit{Sci. Am. (Int. Ed.)\/} \textbf{286} 66 

\bibitem{Dassarma_AS2001} Das Sarma S 2001 \textit{Am. Sci.\/} \textbf{89} 516

\bibitem{Rashba_PRB2000} Rashba E I 2000 \textit{Phys. Rev. B.\/} \textbf{61} R16267

\bibitem{Ohno_SCIENCE1998} Ohno H 1998 \textit{Science\/} \textbf{281} 951

\bibitem{Ohno_JMMM1999} Ohno H 1999 \textit{J. Magn. Magn. Mater.\/} 
\textbf{200} 110

\bibitem{Kawakami_APL2000} Kawakami K, Johnston-Halperin E, Chen L F, Hanson M, Guebels N, Speck J S, Gossard A C, and Awschalom D D 2000 
\textit{Appl. Phys. Lett.\/} \textbf{77} 2379

\bibitem{Dietl_SCIENCE2000} Dietl T, Ohno H, Matsukara F, Cib\'ert J and 
Ferrand D 2000 \textit{Science\/} \textbf{287} 109

\bibitem{Pearton_JAP2003} Pearton S J, Abernathy C R, Overberg M E, Thaler G T,
Norton D P, Theodoropolou N, Hebard A F, Ren F, Kim J and Boatner L A 2003 
\textit{J. Appl. Phys.\/} \textbf{93} 1

\bibitem{Ueda_APL2001} Ueda K, Tabata H and Kawai, T 2001 
\textit{Appl. Phys. Lett.\/} \textbf{79} 988

\bibitem{Kim_PhysicaB2003} Kim J H, Kim, H, Kim D, Ihm Y and Choo W K 2003 
\textit{Physica B\/} \textbf{327} 304

\bibitem{Lim_SSC2003} Lim S W, Hwang D K and Myoung J M 2003 
\textit{Solid State Commun.\/} \textbf{125} 231 

\bibitem{Sharma_NatMat2003} Sharma P, Gupta A, Owens F J, Rao K V, Sharma R, 
Ahuja R, Guillen J M O, Johannson B and Gehring G A 2003 
\textit{Nature Mater.\/} \textbf{2} 673

\bibitem{Yeo_APL2004} Heo H W, Ivill M P, Ip K, Norton D P, Pearton S J, 
Kelly J G, Rairigh R, Hebard A F and Steiner T 2004 
\textit{Appl. Phys. Lett.\/} \textbf{84} 2292

\bibitem{Fukumura_APL2001} Fukumura T, Jin M, Kawasaki T, Shono T, Hasegawa T, 
Koshihara S and Koinuma, H. \textit{Appl. Phys. Lett.\/} \textbf{78} 958

\bibitem{Kolesnik_JSupercond2002} Kolesnik S, Dabrowski B and Mais J 2002 
\textit{J. Supercond.\/} \textbf{15} 251

\bibitem{Risbud_PRB2003} Risbud A S, Spaldin N A, Chen Z Q, Stemmer S and 
Seshadri R 2003 \textit{Phys. Rev. B\/} \textbf{68} 205202

\bibitem{Lawes_XXX2004} Lawes G, Risbud A S, Ramirez A P and Seshadri R 
cond-mat/0403196

\bibitem{Kolesnik_JAP}
Kolesnik S, Dabrowski B,  and Mais J 2004
\textit{J. Appl. Phys.\/} \textbf{95} 2582

\bibitem{Spaldin_PRB2004} Spaldin N A 2004 \textit{Phys. Rev. B\/}
\textbf{69} 125201

\bibitem{Chiba_JSSC1997} Chiba H, Atou T and Syono Y 1997 
\textit{J. Solid State Chem.\/} \textbf{132} 139

\bibitem{Kohn_JSSC1976} Kohn K, Inoue K, Horie O and Akimoto S-I 1976 
\textit{J. Solid State Chem.\/} \textbf{18} 27

\bibitem{Garrett_MRB1981} Garrett J D, Greedan J E and MacLean D A 1981 
\textit{Mater. Res. Bull.\/} \textbf{16} 145

\bibitem{Matthias_PRL1961} Matthias B T, Bozorth R M and van Vleck J H 1961 
\textit{Phys. Rev. Lett.\/} \textbf{7} 160

\bibitem{Sampath_JACerS1998} Sampath S K and Cordaro F 1998 
\textit{J. Am. Ceram. Soc.\/} \textbf{81} 649

\bibitem{Andeen_JCrystG2003} Andeen D, Loeffler L, Padture N and 
Lange F F 2003 \textit{J. Crys. Growth\/} \textbf{259} 103

\bibitem{Loeffler_JMR2004} Loeffler L and Lange F F 2004 
\textit{J. Mater. Res.\/} \textbf{19} 902

\bibitem{EFFINO} Spiering H, Deak L, Bottyan L 2000
\textit{Hyperfine Interactions\/} \textbf{125} 197.

\bibitem{Berar_xnd} B\'erar J F and Garnier P 1992 computer code XND available 
from the website at http://www.ccp14.ac.uk

\bibitem{Josties_NJMM1995} Josties M, O'Neill H S C, Bente K and Brey G 1995 
\textit{Neus Jahr. Mineral. Monat.\/} 273

\bibitem{Fleet_ActaCB1982} Fleet M E 1982 \textit{Acta Crystallogr. B\/} 
\textbf{38} 1718

\bibitem{Sundman_JPhaseEq1991} Sundman B 1991 \textit{J. Phase Equil.\/} 127

\bibitem{Smirnov_EPSL2002} Smirnov A V and Tarduno J A 2002 
\textit{Earth Planet. Sci. Lett.\/} \textbf{194} 359

\end{thebibliography}
\end{document}